\documentclass[twocolumn]{aastex631}

\newcommand{\Msun}{$\mathrm{M}_\odot$}
\newcommand{\Rsun}{$\mathrm{R}_\odot$}

\newcommand{\appropto}{\mathrel{\vcenter{
  \offinterlineskip\halign{\hfil$##$\cr
    \propto\cr\noalign{\kern2pt}\sim\cr\noalign{\kern-2pt}}}}}

\begin{document}

\title{The circumstellar material around the Type IIP SN 2021yja}

\correspondingauthor{Alexandra Kozyreva}
\email{sasha@mpa-garching.mpg.de}

\author[0000-0001-9598-8821]{Alexandra Kozyreva}
\affiliation{Max-Planck-Institut f\"ur Astrophysik, Karl-Schwarzschild-Str. 1, 85748 Garching bei M\"{u}nchen, Germany}

\author{Jakub Klencki}
\affiliation{European Southern Observatory, Karl-Schwarzschild-Strasse 2, 85748 Garching bei M\"{u}nchen, Germany}

\author{Alexei V. Filippenko}
\affiliation{Department of Astronomy, University of California, Berkeley, CA 94720-3411, USA}

\author{Petr Baklanov}
\affiliation{NRC ``Kurchatov Institute'' -- ITEP, Moscow, 117218, Russia}
\affiliation{Keldysh Institute of Applied Mathematics, Russian Academy of Science, Miusskaya sq. 4, 125047 Moscow, Russia}

\author{Alexey Mironov}
\affiliation{Sternberg Astronomical Institute of Lomonosov Moscow State University, 119992, Moscow, Russia}

\author{Stephen Justham}
\affiliation{Anton Pannekoek Institute of Astronomy and GRAPPA, Science Park 904, University of Amsterdam, 1098XH Amsterdam, The Netherlands}
\affiliation{School of Astronomy \& Space Science, University of the Chinese Academy of Sciences, Beijing 100012, People's Republic of China}
\affiliation{Max-Planck-Institut f\"ur Astrophysik, Karl-Schwarzschild-Str. 1, 85748 Garching bei M\"{u}nchen, Germany}

\author{Andrea Chiavassa}
\affiliation{Universit\'{e} C\^{o}te d'Azur, Observatoire de la C\^{o}te d'Azur, CNRS, Lagrange, CS 34229, Nice, France}
\affiliation{Max-Planck-Institut f\"ur Astrophysik, Karl-Schwarzschild-Str. 1, 85748 Garching bei M\"{u}nchen, Germany}

\begin{abstract}
The majority of Type II-plateau supernovae (SNe~IIP) have light curves that
are not compatible with the explosions of
stars in a vacuum; instead, the light curves require the progenitors to be
embedded in circumstellar matter (CSM). 
We report on the successful fitting of the well-observed SN~IIP
2021yja as a core-collapse explosion of a massive star
with an initial mass of $\sim 15$~\Msun\ and a pre-explosion radius of 631~\Rsun.
To explain the early-time behaviour of the broad-band light curves, the presence of
0.55\,\Msun\ CSM within $\sim 2 \times 10^{\,14}$\,cm is needed. Like many other
SNe~IIP, SN~2021yja exhibits an early-time flux excess including ultraviolet
wavelengths. This, together with the short rise time ($< 2$\,days) in the $gri$
bands, indicates the presence of a compact component in the CSM,
essentially adjacent to the progenitor.
We discuss the origin of the
pre-existing CSM, which is most likely a common property of highly
convective red supergiant envelopes. 
We argue that the difficulty in fitting the
entire light curve with one spherical distribution indicates that the CSM
around the SN~2021yja progenitor was asymmetric.

\end{abstract}


\section[Introduction]{Introduction}
\label{sect:intro}

The evidence for exploding massive stars being surrounded
by circumstellar matter (CSM) increases with almost each newly discovered Type
II-plateau supernova
\citep[SN~IIP;][]{2017NatPh..13..510Y,2018NatAs...2..808F,2018ApJ...858...15M,2019AandA...625A...9D,2020ApJ...895L..45G,2021NatAs...5..903H,2021ApJ...912...46B}.
Up to 70\% of SNe\,IIP cannot be explained by progenitors located in
a vacuum, but require the presence of nearby CSM. 
The CSM may be
explained by an extraordinarily strong, steady wind that a massive
star experiences during the few years prior to exploding \citep{2014ARAandA..52..487S}.
Just before collapse, core silicon
burning lasts only a day, and core oxygen burning lasts a year. 
One could speculate that the claimed wind happens during
the last years of core carbon burning and core neon burning. 
Other possible mechanisms for producing this expulsion of the
outermost layers include the $\varepsilon$-mechanism
\citep{2020ApJ...891L..32M,2022MNRAS.512.2777T} or gravity/acoustic waves \citep{2017MNRAS.470.1642F}.
As we discuss here, yet another option is that the binary undergoes a
common-envelope (CE) phase, merges, and ejects a
fraction of the CE considered for dense CSM around some extreme SNe by
\citet{2012ApJ...752L...2C}.
However, we argue that the most natural scenario for SN~2021yja and similar
IIP SNe is a globally asymmetric structure of the red
supergiant (RSG) and its tenuous atmosphere caused by convective motions.

The recently discovered SN~IIP 2021yja \citep{2022arXiv220308001V,2022arXiv220308155H}
shows a blue excess in its spectra at early times.
\citet{2022arXiv220308155H} claim a wind mass-loss rate of
$10^{-6}$\,\Msun\,yr$^{-1}$, normal for 
RSG winds
\citep[][]{2017MNRAS.465..403G,2020MNRAS.492.5994B}\footnote{The whole range
of stellar winds during the RSG phase is\\
$\dot{M}=10^{\,-7}$--$10^{\,-4}$\,\Msun\,yr$^{\,-1}$ \citep{2010A&A...523A..18D}.}. In this
study we test the hypothesis of
the possible presence of matter surrounding SN~2021yja. 
The reconstructed colour-temperature evolution during the first 20\,days shows
the decline of $T_\mathrm{col} \propto t^{-0.6}$ predicted by shock-cooling models
\citep{2010ApJ...725..904N,2011ApJ...728...63R,Shussman2016,2019ApJ...884...41F}.
As suggested by \citet{2020MNRAS.494.3927K}, the colour temperature during this phase
is a good indicator of the progenitor radius; thus,
\citet{2022arXiv220308155H} approximate it to be 900\,\Rsun\ or
2000\,\Rsun, depending on the assumed shock-cooling model, respectively
\citet{Shussman2016} and \citet{2017ApJ...838..130S}.
As correctly mentioned by \citet{2022arXiv220706179M}, an analytic formulation
always provides a very rough estimate and serves as an approximate diagnostic. 
In fact, \citet{2022arXiv220308155H}
compare the colour-temperature evolution of SN~2021yja to a set of numerical
simulations of RSG models. The numerical
simulations agree well with the analytic formulation by
\citet{Shussman2016}, who present updated versions of the formulae by \citet{2010ApJ...725..904N}.
According to \citet{Shussman2016}, $T_\mathrm{col} \propto
R^{\,0.46}\,E^{\,-0.25}$, where $R$ is the progenitor radius and $E$ is an explosion energy.
For simplicity, the energy is dropped off in the mentioned comparison,
while $E$ is close to 1~foe (1\,foe $= 10^{51}$\,erg) and the dependence on
$E$ is weaker than dependence on $R$. 
If the explosion energy of SN~2021yja differs from 1~foe,
the progenitor radius estimate varies accordingly as $R \propto E^{\,0.25/0.46}
\approx E^{\,0.54}$ for the same colour temperature. Hence, if the energy is
50\% higher then the radius is larger by a factor of 1.3, although it is
still a rough estimate.
Nevertheless, we emphasise that the inferred CSM/wind is optically thin
before the shock propagation and does not affect the estimated progenitor radius
\citep{2017AandA...605A..83D}. 

The paper is structured as follows. In Section~\ref{sect:method} we describe
our two best-fitting models. Section~\ref{sect:results} presents our broad-band
light curves, which match the observed light curves of SN~2021yja, are
presented in Section~\ref{sect:results},
and in Section~\ref{sect:conclusions} we summarise our findings.

\section[Input models]{Input model and Method}
\label{sect:method}

We used model m15 from \citet{2019MNRAS.483.1211K} --- namely, the case with a high
explosion energy of 1.53\,foe. This is a 15\,\Msun\
solar metallicity stellar evolution model computed with \verb|MESA| \citep{2015ApJS..220...15P}
and exploded with \verb|V1D| \citep{1993ApJ...412..634L}. We consider two values
for the total mass of radioactive nickel: 0.175\,\Msun\ (model m15ni175) and
0.2\,\Msun\ (model m15ni2), although \citet{2022arXiv220308155H} and
\citet{2022arXiv220308001V} claim a $^{56}$Ni mass
of 0.141\,\Msun\ and 0.2\,\Msun, respectively, based on the radioactive tail luminosity. 
The $^{56}$Ni mass fraction was scaled to have a total mass of 0.175\,\Msun\ or
0.2\,\Msun\ while reducing the mass fraction of silicon. We set the higher
mass of $^{56}$Ni because the tail luminosity is not matched by a model with 0.141\,\Msun\
of radioactive nickel. 0.2\,\Msun\ of $^{56}$Ni is at the upper limit of the
total amount of radioactive nickel produced in a neutrino-driven explosion;
however, it is still within the range of accepted uncertainties
\citep{2016ApJ...818..124E,2016ApJ...821...38S,2020ApJ...890...51E}.
Moreover, if there is any asymmetry in the SN ejecta, the effective  
$4\pi$-equivalent mass of $^{56}$Ni might be higher
\citep{2022MNRAS.514.4173K,2022A&A...657A..64S}.

\begin{figure}
\centering
\vspace{4mm}\hspace{-7mm}\includegraphics[width=0.5\textwidth]{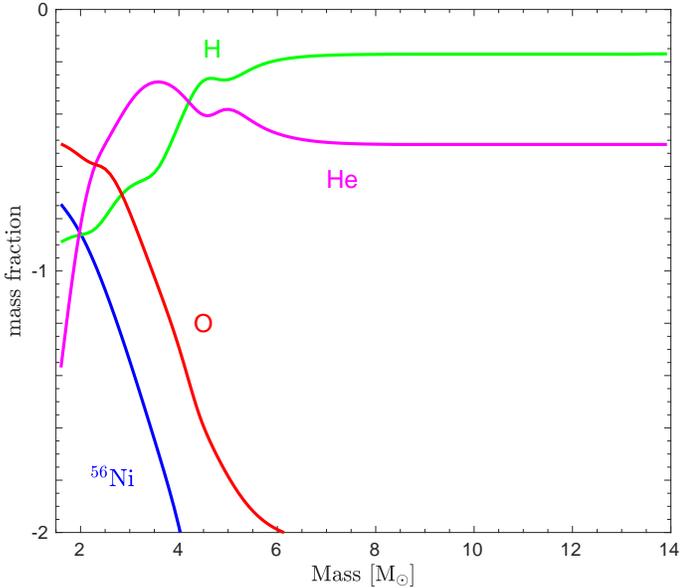}
\caption{Chemical structure of model m15ni175.
We show selected elements: hydrogen (green), helium (magenta), oxygen
(red), and radioactive nickel (blue).}
\label{figure:chemie}
\end{figure}

We notice that the shape of the transition from the plateau to
the radioactive tail in the bolometric light curve of SN~2021yja is very shallow. This
is similar to the light curve of a self-consistent explosion of a 15\,\Msun\ progenitor
and its three-dimensional (3D) post-explosion hydrodynamics simulations carried out
with the \verb|PROMETHEUS-VERTEX| code
\citep{2015AandA...577A..48W,2017ApJ...846...37U}. In these simulations the SN
ejecta undergo strong macroscopic mixing. The iron-group
elements, including radioactive nickel, are mixed far beyond the core
and penetrate the hydrogen-rich envelope. Conversely, hydrogen is mixed deep
into the interior of the SN ejecta. The combination of radioactive nickel
mixed thoroughly with hydrogen in the ejecta leads to a shallow
and smooth drop from the
plateau. Therefore, in our current study we modify model m15 by artificially
mixing hydrogen inward. The final chemical composition of model
m15ni175 is shown in Figure~\ref{figure:chemie}.

As SN~2021yja has a blue excess at early times, \citet{2022arXiv220308155H}
conclude that the progenitor exploded in a pre-existing tenuous
environment which the authors call a ``weak wind.'' This particular wind rate was
chosen based on the synthetic observables computed by \citet{2017AandA...605A..83D}.
Therefore, we assume the existence of matter around the
progenitor. We carried
out simulations for a variety of radii, ``interface'' density of the CSM (the
density where the CSM is adjacent to the progenitor), density slopes, and CSM
with shells. Here we
report only on the most successful models in which the CSM was directly attached
to the surface of the progenitor. The density slope was assumed to conform to $\rho \propto
r^{-2}$, and the extent of the CSM was chosen to be
$1.8 \times 10^{\,14}$\,cm (2700\,\Rsun).  In total, the CSM mass is
0.55\,\Msun. Assuming a wind origin for the CSM, a rough estimate of the wind
mass-loss rate is
\begin{equation}
  \dot{M} \approx {\Delta M_\mathrm{wind} v\over{R}}=
      {(0.55\,\mathrm{M}_\odot)(20\,\mathrm{km\,s}^{-1})
        \over{1.8 \times 10^{14}\,\mathrm{cm}}} \approx
      0.18\,\mathrm{M}_\odot\,\mathrm{yr}^{-1}\,,
\label{equation:wind}
\end{equation}
where we input a typical RSG wind velocity of 20\,km\,s$^{-1}$ \citep{2017MNRAS.465..403G,2018MNRAS.475...55B}.
We discuss the proposed mass of the CSM material in Section~\ref{sect:origin}.
Eight additional models with a variety of CSM configurations are presented
in Appendix~\ref{appendix:append1}.

\begin{figure}
\centering
\vspace{4mm}\hspace{-7mm}\includegraphics[width=0.5\textwidth]{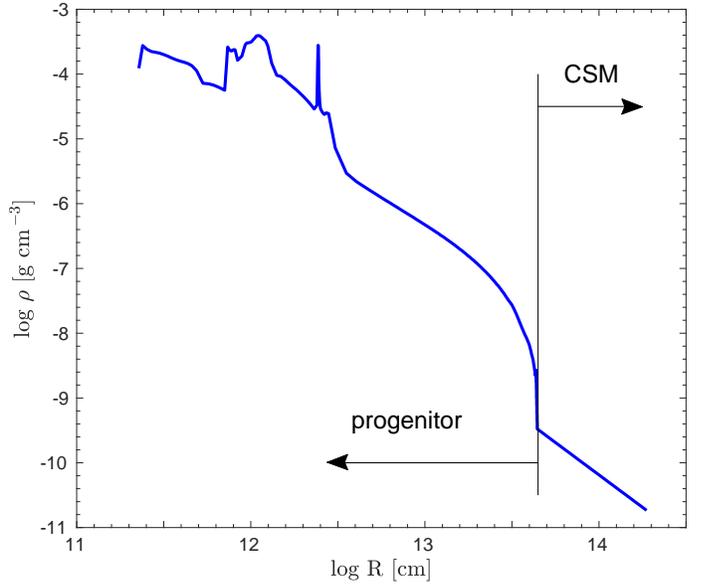}
\caption{Density structure of model m15ni175.}
\label{figure:rho}
\end{figure}

Models m15ni175 and m15ni2 were mapped into the 1D radiation-hydrodynamics
code \verb|STELLA| \citep{2006AandA...453..229B}
\footnote{The version of STELLA used here is private, not the one
implemented in MESA \citep{2018ApJS..234...34P}.}.  \verb|STELLA| is
capable of processing hydrodynamics, including shock propagation and its
interaction with the medium, as well as the radiation field
evolution --- computing light curves, spectral energy distributions, and the resulting
broad-band magnitudes and colours.  We use the standard parameter settings,
well-explained in many papers involving \verb|STELLA| simulations \citep[see,
e.g.,][]{2021AstL...47..291T,2020MNRAS.497.1619M}.  The
thermalisation parameter is set to 0.9, as
recommended by the recent study of \citet{2020MNRAS.499.4312K}.

We note that the CSM is added as an attached density profile with the
same temperature and chemical composition as the last zone of the progenitor model,
and zero velocity artificially. Therefore, the stellar structure with the attached CSM is not
in hydrodynamical and thermal equilibrium. Consequently, the
radiation field is not fully trustable during roughly the first day. In the
plot showing the rising part of the light curve in Section~\ref{sect:rise},
we deliberately do not show the first day of simulations.

\section[Results]{Results}
\label{sect:results}

\subsection[Light curve]{Light curve}
\label{subsect:lc}

\begin{figure}
\centering
\vspace{2mm}
\includegraphics[width=0.5\textwidth]{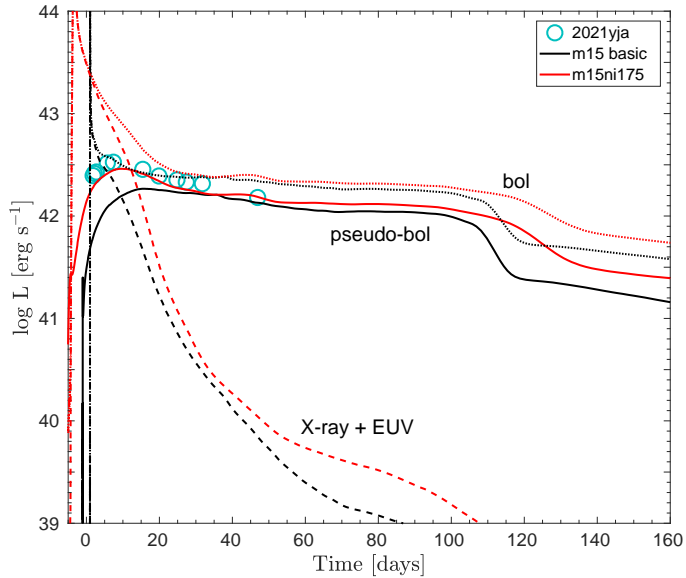}
\caption{Pseudobolometric (solid), bolometric (dotted), and X-ray-EUV (dashed),
light curves for model m15ni175 (red), along with the ``bolometric'' data of SN~2021yja (circles). 
We show the basic model m15 in black for reference and to demonstrate the effect of interaction.}
\label{figure:bol}
\end{figure}

\begin{figure*}
\centering
\vspace{9mm}
\includegraphics[width=\textwidth]{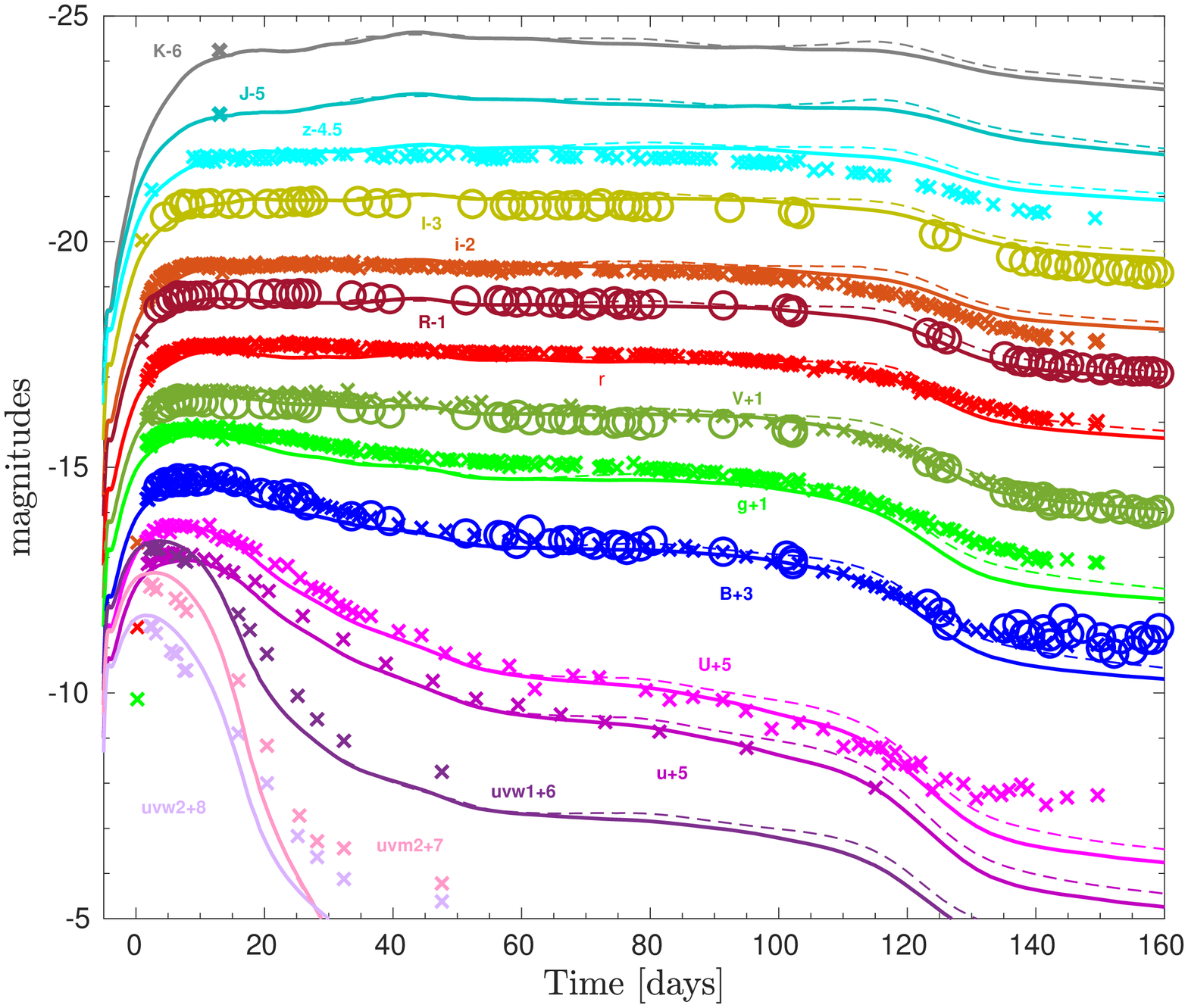}
\caption{Broad-band light curves for models m15ni175 (thick curves) and m15ni2 (dashed curves), together with
SN~2021yja. Crosses represent the observations taken from \citet{2022arXiv220308155H}, and
circles represent data taken from \citet{2022arXiv220308001V}. The statistical uncertainties
of the observations are smaller than the data points. Time ``0''
corresponds to the explosion epoch introduced by \citet{2022arXiv220308155H}.
The synthetic curves are shifted by $-8$\,days to match the early phase and
the end of the plateau.}
\label{figure:bands}
\end{figure*}

In Figure~\ref{figure:bol}, we present pseudobolometric, bolometric, and
X-ray-EUV ($\lambda<325$\,\AA) 
light curves for the models m15ni175 (which we call ``best-fit'' or ``best'' hereafter), 
as well as the basic model m15, with
the ``bolometric'' data of SN~2021yja superposed. We note that the synthetic
bolometric light curves of our models are truly bolometric, while the 
``bolometric'' light curve of SN~2021yja is not truly bolometric; instead, it is
constructed based on photometric data fitted to a blackbody via
Markov-Chain-Monte-Carlo light-curve fitting \citep{2020zndo...4312178H}.
In this study, we intend to compare our synthetic observables to the
broad-band magnitudes to avoid misinterpretation while comparing to
the derived bolometric light curve of SN~2021yja.
Figure~\ref{figure:bol} shows the default progenitor m15 from \citet{2019MNRAS.483.1211K}
as a reference to illustrate the effect of the CSM.
The shock propagates through the extended surrounding medium and ionises it,
turning it into a hot, expanding, relatively optically thick layer. Later cooling in
this tenuous layer is responsible for the high luminosity in bluer bands.
We avoid calling this mechanism ``interaction'' since this is the natural
propagation of the shock in the pre-existing CSM adjacent to the progenitor. 
The difference between the default model m15 and model m15ni175,
clearly demonstrates that the presence of an extended medium plays a significant
role in shaping the early-time light curve during the first 35 days.

Figure~\ref{figure:bands} shows the broad-band light curves for our models
m15ni175 and m15ni2, together with photometric data SN~2021yja taken from \citet{2022arXiv220308155H}
and \citet{2022arXiv220308001V}. We note that these two observational
studies propose two different distance moduli and extinction values, even though
the derived absolute magnitudes do not differ significantly.  The time ``0'' is the same as in
\citet{2022arXiv220308155H} --- the estimated explosion epoch, which
differs from the actual moment of core collapse. 
The explosion epoch
introduced by \citet{2022arXiv220308155H} is calculated based on the
nondetection, the first detection, and the assumed approximation. In
nature, the shock breaks out
on the progenitor's surface $\sim 1$\,day after the actual explosion, if the
progenitor is located in a vacuum (for the basic model), and becomes ``visible''.
In the case when the progenitor is embedded in the
CSM (e.g., in the best-fit model), the shock breaks out after 4~days at the edge of the CSM. 
Hence, the explosion epoch in Figure~\ref{figure:bands} is not the same as the actual
moment of core collapse, because there is a time
lapse until the explosion becomes ``visible'', and there is some relative degree of
freedom to set the time shift for the synthetic curves.
The light curves of SN~2021yaj in the majority of broad bands ($U\!V\!W\!1$,
$U\!V\!M\!2$, $U\!V\!W\!2$, $uU\!BgV\!rRiIzJK$)
are matched by our
synthetic light curves well during the entire observed period. We show 
models with two different masses of radioactive nickel, since in some
broad bands on the tail the model with 0.2~\Msun{} of nickel fits the data better. There is
some disagreement in the $U$ and $z$ bands at later times, after day~130, when the ejecta
become more transparent, and proper spectral synthesis is required.

Our best-fit models do not match the first data points collected in the $gri$ bands
at day 0.225 with the MuSCAT3 instrument. We discuss a possible solution for
this tension in Section~\ref{sect:rise} while introducing a model with
the CSM having a different density profile. 
The synthetic light curves in the UV bands $U\!V\!W\!1$, $U\!V\!M\!2$, and $U\!V\!W\!2$
also reproduce the observed magnitudes reasonably well, although they
slightly overestimate the flux during the first 10\,days, and underestimate the flux
after day\,20. The same model introduced to match the first $gri$ points with the different CSM density gradient
fits the UV bands better for the earlier epoch.
Hence, we assume that SN~2021yja
might have CSM with asymmetric density structure.

\subsection[Photospheric velocity]{Photospheric velocity}
\label{subsect:velo}

\begin{figure}
\centering
\vspace{4mm}
\includegraphics[width=0.5\textwidth]{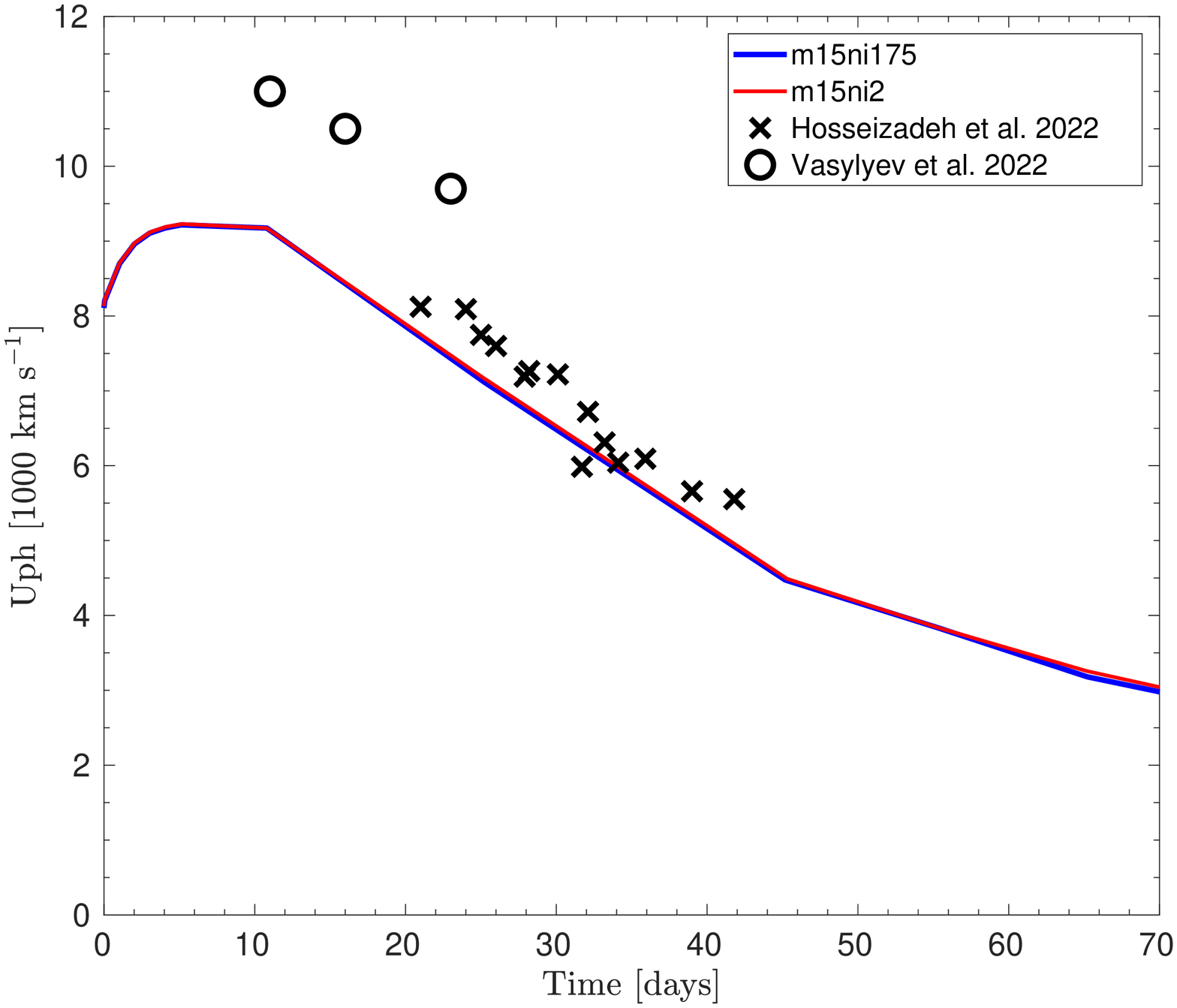}
\caption{Photospheric velocity evolution for models m15ni175 and m15ni2, and for
SN~2021yja. Crosses represent velocities obtained from fits to the iron line in \citet{2022arXiv220308155H}, and circles
represent velocities from spectral modelling in \citet{2022arXiv220308001V}.}
\label{figure:uph}
\end{figure}

In Figure~\ref{figure:uph}, we present the synthetic photospheric velocity evolution
derived from the location of the photosphere in the $B$ band (where the integrated optical
depth in $B$ equals 2/3) and the photospheric velocity evolution of
SN~2021yja. The data taken from \citet{2022arXiv220308001V} are derived via
spectral modelling, while the data taken from \citet{2022arXiv220308155H}
are derived via the absorption minimum of the P~Cygni profile of spectral line
Fe\,II $\lambda5169$. The agreement between our prediction and the velocity estimate
from the iron line is very satisfactory. However, our models do not explain the velocity
estimate derived from the spectral modelling, though this is to some extent model-dependent
and should be considered with caution.
For example, there is the epoch at day 20--21 when the photospheric velocity is estimated by both
methods --- TARDIS spectral modelling and via the iron line. The resulting values are
different: 9700\,km\,s$^{-1}$ and 8100\,km\,s$^{-1}$, respectively.
The photospheric velocities estimated via TARDIS
modelling are systematically overestimate by 10--20\% relative to those
calculated via Fe or Sc lines.
The estimates for SN~2005cs and SN~1999em \citep[Figure~11,][]{2022arXiv220308001V}
give 4000--4500\,km\,s$^{-1}$ (day 15) and 6000\,km\,s$^{-1}$ (day 30),
respectively, while using TARDIS spectral modelling methodology.  
On the contrary, \citet{2009MNRAS.394.2266P} estimate velocities of
3400--3800\,km\,s$^{-1}$ for SN~2005cs and 5500\,km\,s$^{-1}$ for SN~1999em at the same
epochs via spectral lines.

\vspace{1mm}
Overall, the broad-band light curves and photospheric-velocity evolution are
matched reasonably well with our modified 15\,\Msun\ stellar model embedded
in confined CSM. This agrees with findings by
\citet{2022arXiv220308001V}, who
reported a possible 15\,\Msun\ progenitors based on the analysis of archival
{\it Hubble Space Telescope} data.

\section[Possible origin of the CSM]{Possible origin of the CSM}
\label{sect:origin}

The amount of CSM required by matching the light curves of SN~2021yja is 0.55\,\Msun.
Assuming that the CSM was produced by a steady wind, the corresponding wind
estimate is 0.18\,\Msun\,yr$^{-1}$ (see Eq.~\ref{equation:wind}), 
which is extraordinarily high and cannot be explained by a normal, steady, RSG wind mass-loss rate. 
However, the mass of CSM defined by our simulations is in good
agreement with estimates done for other SNe\,IIP \citep{2018ApJ...858...15M,2020ApJ...895L..45G}. 
Below we discuss a few possible origins for the pre-existing CSM.

%\subsection[CE ejection]{CE ejection}
\subsection{Binary interaction} 
\label{subsect:cee}

Let us consider that the progenitor is a
member of a binary system and that the CSM originates from recent
mass-transfer interaction. The best-fitting model requires that $\sim
0.55$\,\Msun\ of matter is ejected on a short timescale of just a few years.  This
is much more than could be lost from a binary as a result of any stable
(nondynamical) mass transfer; a typical rate of thermal-timescale
mass transfer is $\sim 10^{-3}$--$10^{-2}$\,\Msun\,yr$^{-1}$.  Instead, a feasible
possibility is mass transfer that becomes dynamically unstable, leading to 
CE inspiral and a stellar merger, during which a portion of the RSG
envelope may become unbound
\citep[e.g.,][]{Podsiadlowski2001,Morris2006,Ivanova2020}. The onset of the CE
would have to be rapid in order to prevent significant pre-CE mass loss from
the L2 and L3 points \citep{Pejcha2014,Pejcha2017,MacLeod2020,Blagorodnova2021} that
would extend the CSM to larger radii ($\sim 10^{15}$--$10^{16}$\,cm). 
This could be achieved through Darwin instability \citep{Darwin1879},
granted that the companion is $\gtrsim 6$ times less massive than the RSG
\citep[Fig.~8 in][assuming $\eta_1 \approx 0.15$ typical for RSGs]{MacLeod2017}.
Given the low binding energies of RSG envelopes
\citep{Klencki2021}, even such a low-mass companion could generate enough
energy during a CE inspiral to eject $\gtrsim 0.5$\,\Msun{} of matter.  This
scenario, although in principle viable, is however extremely rare, as it
would require the mass-transfer interaction to occur just a few years
before core collapse.  In Appendix~\ref{appendix:append1}, based on 1D stellar models
of radial expansion of RSGs, we estimate the rate of such events as $\lesssim 10^{-4}$
of all SNe, occurring preferentially at very low metallicities of a few
percent of $Z_\odot${}. 
We note that mass transfer close to core collapse is much more
likely in the case of Type Ib or Ic SNe, particularly at low metallicity, 
where the previously (partially) stripped helium star is expanding during 
advanced burning, leading to another
interaction \citep{2002MNRAS.331.1027D,Laplace2020,2022A&A...662A..56K}.

A more likely signature of a recent mass-transfer event is extended CSM
at a radial distance $\sim 10$--100 times the size of the primary.  Unstable
but also stable and nonconservative mass transfer can lead to slow
outflows of mass from the system, most likely concentrated in the equatorial
plane.  Moving at $\sim 20$\,km\,s$^{-1}$, such slow ejecta would need $\sim
200$\,yr to reach $10^{16}$\,cm, making the time window for the onset of mass
transfer much wider.  Mass-transfer events occurring thousands of years
prior to core collapse may still contribute to the CSM at the moment of
explosion in the form of a circumbinary disk that remains from the original
interaction \citep{Kashi2011,Pejcha2016}.  Motivated by this, in
Appendix~\ref{appendix:append2} we test additional progenitor models with a
more-extended CSM and shells. In particular,
we constructed models with the CSM extending to radii of up to
143,000\,\Rsun\ ($10^{16}$\,cm) and having a mass of 0.05\,Msun, as well 
as a model with the same CSM profile and a 0.16\,\Msun\ shell inserted at
the edge of the CSM. The results of
radiative-transfer simulations for these cases are
presented in Appendix~\ref{appendix:append2} together with other attempts. 
The light curve for the model
with the shell at a distance of 143,000\,\Rsun\ has two distinct maxima,
which are not observed in SN~2021yja. Moreover, the actual interaction leads
to distinct spectroscopic signatures --- narrow lines, which are also not observed in SN~2021yja.
Therefore, this experiment illustrates that the observables of SN~2021yja
cannot be reproduced by a model with a shell; thus,
binary interaction is unlikely to have played a role in
shaping the CSM around SN 2021yja.

\subsection[Convective nature of the RSG atmospheres]{Convective nature of the RSG atmospheres}
\label{subsect:convection}

Studies by \citet{2009A&A...506.1351C}, \citet{2011A&A...535A..22C}, \citet{2022ApJ...929..156G}
and others show that an RSG has a convective envelope,
with a characteristic size of the convective cell on the order of the size of the star itself
(up to $\sim 1000$\,\Rsun; e.g., \citealt{1999AJ....117..521V, 2013MNRAS.434..437C,
  2014A&A...566A..88A}). 
Convection is inferred from giant structures observed at the stellar
surface, with sizes comparable to the stellar radius and evolving on weekly
or yearly timescales \citep[e.g.,][]{2010A&A...515A..12C,2011A&A...528A.120C,2018A&A...614A..12M,2018Natur.553..310P}.
This results in extreme atmospheric conditions with large variations in velocity,
density, and temperature producing strong radiative shocks in their extended
atmosphere that can cause the gas to levitate and thus contribute to
mass loss \citep{2017A&A...600A.137F,2011A&A...535A..22C}.
%The convective motion supports macroscopic mixing of matter
%and leads to expulsion of high-entropy material from the surface of the star. 
We note that the stellar radius is defined as a surface where the integrated Rosseland mean 
depth equals unity. This means that there is some amount of gravitationally
bound stellar matter beyond the optical depth of unity
\citep{2019A&A...632A..28K}, which can be swept to a distance of 1500\,\Rsun\ or more.
The asymmetry of extended material is not a unique property of an RSG.
Oblate extended atmospheres were observed in nearby RSGs such as
VX\,Sagittarii \citep{2010A&A...511A..51C,2022AandA...658A.185C}, 
Betelgeuse \citep{2009A&A...508..923H,2021Natur.594..365M},
CE Tauri \citep{2018A&A...614A..12M},
AZ Cygni \citep{2021ApJ...919..124N},
V602 Carinae \citep{2020A&A...635A.160C}, 
V766 Centauri \citep{2017A&A...606L...1W}, 
VY CMa \citep{2019A&A...627A.114K}, and 
yellow hypergiant IRC$+$10420 \citep{2022arXiv220705812K}.
This kind of star being
part of a close binary is likely to change its shape according to the
equipotential surfaces (the Roche-lobe surface is the critical equipotential surface
having the Lagrange point L1), which breaks spherical symmetry. 

Therefore, we conclude that our best-fit model which matches the observational
properties of SN~2021yja might be explained by a normal RSG star with an
asymmetric, extended, convective envelope which is also most likely part of an
interacting binary system.
It might also be possible that the high-entropy plume was expelled during
the last stages of evolution (e.g., hundreds to thousands of years prior the
core collapse) and was pointed at the direction close to the line of sight
of the Earth. The probable asymmetry is also
supported by spectropolarimetric observations of SN~2021yja, which suggest a
noticable degree of polarization (Vasylyev et al., in preparation).

\section[The constraint from the rise time as a signature of asymmetry]{The constraint from the rise time as a signature of asymmetry}
\label{sect:rise}

\begin{figure}
\centering
\vspace{5mm}\includegraphics[width=0.5\textwidth]{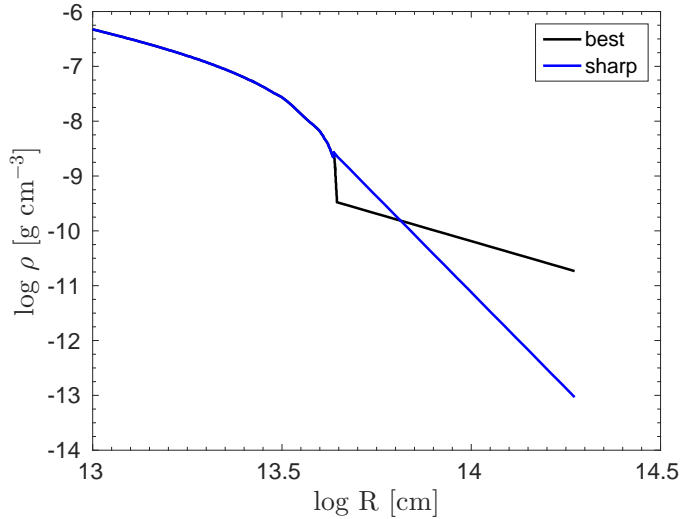}
\caption{Density structures of the best-fit (``best'') model and the model
with the sharper density gradient in the CSM (``sharp'').}
\label{figure:addRiseRho}
\end{figure}

\begin{figure}
\centering
\hspace{-6mm}\includegraphics[width=0.5\textwidth]{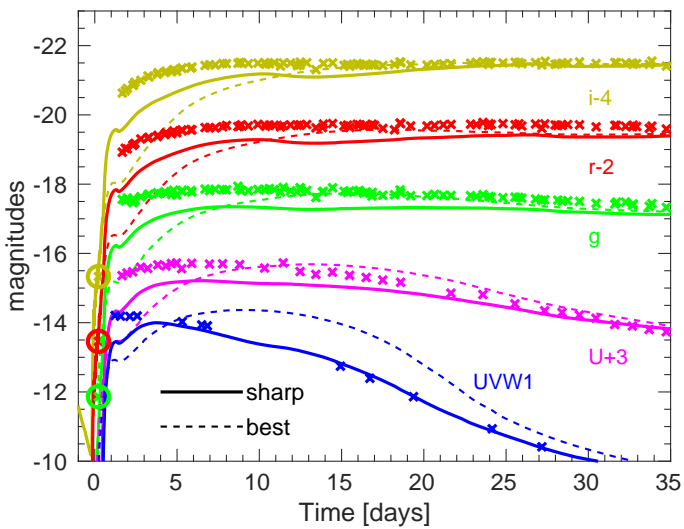}
\caption{$U\!V\!W\!1$ and $U\!gri$ broad-band magnitudes for the best-fit model ``best''
(dashed curves) and the model ``sharp'' with the compact CSM (solid curves),
and SN~2021yja.}
\label{figure:addRiseLCs}
\end{figure}

The high-cadence observations by \citet{2022arXiv220308155H} include
a nondetection limit up to $-3$\,hr before the estimated explosion
epoch and the first detection at day 0.225. Together with the subsequent
observations, SN~2021yja shows a feature common to many SNe~IIP: a very
sharp rise to maximum brightness \citep[e.g.,][]{2015MNRAS.451.2212G}. 
The data in the $gri$ bands are incompatible with our best-fit model, which
overestimates the flux at this epoch. In Figure~\ref{figure:bands},
the synthetic light curves are shifted to match the $U$-band maximum, the
rising phase in the $BV\!g$ bands, and the transition from the plateau to
the tail, although the early-time data in the $gri$ bands are not matched
(exhibiting a sharp rise of 5.5\,mag between day 0.225 and day 1.7).
Therefore, we consider an additional model to match these data points. 
We built a model ``sharp'' with a sharper density gradient, with slope
$\alpha = \Delta\,{\rm log}\,\rho/\Delta\,{\rm log}\,r = -7$
versus $\alpha = -2$ in the best-fit model, and the same CSM extension, 
$R_\mathrm{out} = 1.8 \times 10^{\,14}$\,cm (2700\,\Rsun; see
Fig.~\ref{figure:addRiseRho}). The mass of the CSM in the model ``sharp'' is
0.26\,\Msun{}. The radiative-transfer simulations for this
model result in light curves which have shorter rise times, as seen in
Figure~\ref{figure:addRiseLCs}. We present $gri$ light curves for the
``sharp'' model (together with the ``best'' model) in Figure~\ref{figure:addRiseLCs} 
which match the first detection
data points. Additionally, we show the light curve in the $U$ broad band and 
in the $U\!V\!W\!1$ band. The $U\!V\!W\!1$ light curve of the ``sharp'' model fits the {\it Swift}
data better than the ``best'' model, even though it slightly underestimates the
flux during the first two days, similarly to other bands.

To match the first data points, the best-model ``best'' and the model ``sharp'' with the
sharper CSM have different shifts in time, a 1.6\,day difference.
We note that in Figure~\ref{figure:addRiseLCs} the best-fit model is shifted
by 2.8\,days, while in Figure~\ref{figure:bands} the light curves are shifted
by 8\,days relative to the explosion in the simulations. 
Indeed, the shift in time cannot be arbitrary; we choose the
different shifts to match the data in broad bands. 
As discussed in the previous section,
RSG envelopes have a highly convective nature (i.e., macroscopic plumes).
The 3D structure of the modelled RSG envelope \citep{2022ApJ...929..156G}
shows scatter at a fixed radius coordinate within 1--2 orders of magnitude, 
which in turn means a scatter in the sound speed in
various radial directions and different shock-crossing times. The shock
propagates in different radial directions with different speeds and reaches
the corrugated photosphere at different times \citep{2011A&A...535A..22C,2022arXiv220604134G}. %, which varies within a couple of hours.
As both our models are computed with a 1D radiation-transfer code, we do not
account for different density structures of the progenitor envelope;
therefore, there is physically consistent freedom to apply a relative shift
to one of two light curves. Our model ``sharp'' does not perfectly
match the rise, but we show the tendency of the actual
progenitor plus CSM system.

Hence, we suggest the following physical picture underlying SN~2021yja.
The progenitor of SN~2021yja is consistent with our simulations based on 
an initially 15\,\Msun\ stellar model\footnote{We note that the most
influential parameter is not the initial, but the ejecta mass;
the initial mass could vary depending on the stellar-evolution
calculations and the wind mass-loss prescription.} with the convective envelope and
asymmetric CSM caused by the dynamical nature of the convective envelope.
The explosion forms a shock propagating
nonspherically within the envelope and breaking at the surface
within a day or so. The CSM caused by the plume expulsion forms an asymmetric
density structures surrounding the progenitor. The sharp rise time of about 7\,days
is explained by the model with the sharp density gradient --- that is, the first
light comes from the radiation front in the radial direction of the 
compact CSM, while the major fraction of the light curve is then overwhelmed by
radiation from the radial directions where the density decline is shallower. The
fact that the best-fit model (with the shallower CSM) matches the overall
broad-band light curves reasonably well (assuming different time shift; see
Fig.~\ref{figure:bands}) offers support for our interpretation.

\section[Conclusions]{Summary and conclusions}
\label{sect:conclusions}

We simulated and explained the broad-band light curves
and photospheric-velocity evolution of the bright Type IIP SN~2021yja, which
shows an early UV excess. The best-fit models are initially 15\,\Msun\ RSGs with
an admixture of 0.175\,\Msun\ and 0.2\,\Msun\ of radioactive $^{56}$Ni. 
Light-curve modelling demonstrates the necessity of a high degree of mixing in the
post-explosion SN ejecta: outward mixing of radioactive material and inward mixing of
hydrogen, which combine to provide a smooth transition from the plateau to the tail.
The early-time light-curve evolution is explained by the presence of
0.55\,\Msun\ of CSM adjacent to the progenitor. % (2,700~\Rsun{}).
The CSM might originate from an asymmetric, convective mass ejection shortly
before core collapse, pointing toward the observer. The amount of CSM 
and radioactive nickel in the ejecta might be
lower than in our model, after proper accounting for the asymmetry. 

SN~2021yja is another example of an SN~IIP that requires CSM to explain
the behaviour of early-time light curves.
Overall, SNe~IIP constitute a large, diverse group of explosions of RSGs
surrounded by CSM at various distances, some essentially attached to
the star. Analysis of the mass and extension of the CSM in
these events will help provide a better understanding of the evolution
of massive stars.

%\section*{Acknowledgments}

\vspace{6mm}
We thank Avishay Gal-Yam, Christian Vogl, Stefan
Taubenberger, Sergey Blinnikov, Marat Potashov, Sergey Bykov, and Patrick Neunteufel for fruitful discussions, and Griffin
Hosseinzadeh and Sergiy Vasylyev for providing the data for SN~2021yja.
J.K. acknowledges support from an ESO Fellowship.
P.B. is sponsored by grant RFBR 21-52-12032 in his work on the STELLA code development.
A.V.F. received funding from the Christopher R. Redlich Fund, numerous individual donors, and {\it Hubble Space Telescope} grant GO-16178 from the Space Telescope Science Institute (STScI), which is operated by the Association of Universities for Research in Astronomy (AURA), Inc., under NASA contract NAS5-26555.

\section*{Data availability}

The data computed and analysed for the current study are available via the
link
%\url{
\href{https://wwwmpa.mpa-garching.mpg.de/ccsnarchive/data/Kozyreva2022/}{https://wwwmpa.mpa-garching.mpg.de/ccsnarchive/}.

\bibliography{references}{}
\bibliographystyle{aasjournal}

\appendix
\section[Details about Probability of CEE]{Details about Probability of CEE}
\label{appendix:append1}

Here we estimate what fraction of all SN progenitors are primary stars in
binary systems which engaged in a mass-transfer phase just $\Delta t_{\rm
RLOF-CC}$\,yr before core collapse.  Because we are interested in SN
progenitors that explode as RSGs and have most of their envelope retained at
the time of core collapse, we only consider the first-ever mass-transfer event
from the progenitor (more generally, a component of a binary system can undergo
several distinct phases of mass transfer).  This allows us to use single
stellar models from \citet{Klencki2020} to approximate the evolution of the
primary.  We assume that the companion is $q = 1/10$ of the initial
(zero-age) mass of the primary.

\begin{figure}[h]
\centering
\includegraphics[width=0.5\textwidth]{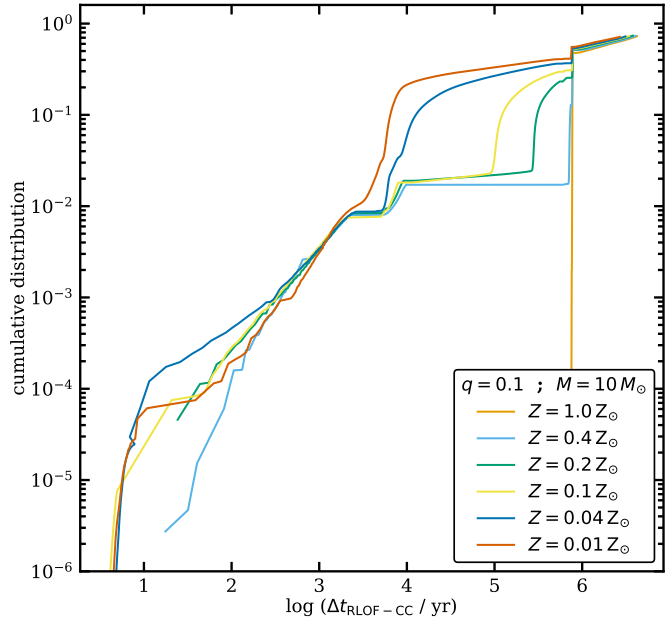}
\caption{Cumulative distribution of the time between the onset of mass transfer and
core collapse, ${\rm log}\,(\Delta t_{\rm RLOF-CC}$\,yr$^{-1})$, estimated for 
a 20\,\Msun\ RSG progenitor evolving in a binary with a 1/10 as massive companion, 
shown for different metallicities. }
\label{fig_tdel_cc}
\end{figure}

For each evolutionary step, we calculate what would need to be the initial
orbital period of the binary for the mass transfer to start at that
particular $\Delta t_{\rm RLOF-CC}$.  Orbital evolution owing to wind mass
loss is taken into account.  We note that an onset of mass transfer is
generally only possible in phases of radial expansion of the primary.  We
convolve the obtained orbital periods with the initial orbital period
distribution of massive binaries, ${\rm d}N/{\rm d \, log} P_{\rm ini}
\propto {\rm log}\,P_{\rm ini}^{-0.55}$ \citep{Sana2012}, normalised to the
range ${\rm log}\,P_{\rm ini} = [0.15, 5.5]$.  This allows us to obtain the
cumulative distribution of ${\rm log}\,(\Delta t_{\rm RLOF-CC}$\,yr$^{-1}$),
shown in Figure~\ref{fig_tdel_cc} for a primary with 20\,\Msun{} and several
different metallicities.  The distribution goes up to $\sim 0.7$ as the
remaining $\sim 30\%$ are members of wide noninteracting binaries.  We
tested that different values of $q$ have negligible effect on the
distribution.  For the range of interest ${\rm log}\,(\Delta t_{\rm RLOF-CC}$\,yr$^{-1}$)
$\lesssim 4$, there is little effect of changing the primary mass as well.

   Figure~\ref{fig_tdel_cc} demonstrates that only about one in every
10,000 SN progenitors is expected to be a primary in a binary system
that had undergone its first phase of mass transfer in the last $\lesssim
10$\,yr.  Figure~\ref{fig_tdel_cc} also gives preference to low-metallicity
progenitors ($Z < 0.1\,{\rm Z}_\odot${}), as the higher-metallicity RSG models from
\citet{Klencki2020} do not expand during the final evolutionary stages owing
to mass loss.  We caution, however, that this result is highly uncertain in
1D stellar models.  In any case, unless a physical mechanism unaccounted for
in the hydrostatic 1D stellar models could cause a significant expansion of
RSGs just years prior to core collapse, it is very unlikely for such a SN
progenitor to experience mass transfer shortly before the explosion.

\section[Additional Models with Different CSM Structures]{Additional models with Different CSM Structures}
\label{appendix:append2}

We computed additional sets of models to test the effect of more extended CSM,
lower density CSM, and CSM in a shell. For this purpose, we constructed
models based on the best-fit progenitor model from the main study, and
attached different modified CSM distributions instead of the CSM profile used as the
best-fit model for SN~2021yja. 

Two classes of profiles represent (1) different kinds of extended
material with a varied CSM density and radius, and (2) CSM with imitated
shells. While the first class might be considered as a steady wind and the
extension of the convective plumes of RSGs, the cases with shells might
be connected to either eruptive wind mass loss, CE ejection, or any kind
of interaction in a close binary system.
%In Figure~\ref{figure:addRhoR}, the models ``best'', ``reduced $\rho$ 1'',
%``reduced~$\rho$~2'', and ``extended 2'' are shown without offset, while
%the models ``extended\,1'', ``extended\,3'', ``Shell\,1'', ``Shell\,2'', and
%``Shell\,3'' are shown with different offsets for clarity. The models
%``extended~2'',  ``extended\,3'', ``Shell\,1'', ``Shell\,2'', and ``Shell\,3'' have 
%the same ``attached'' density (see details in Table~\ref{table:addModels}).
%For the models with shells, the increased density is inserted on top of it locally. 

\begin{figure}
\centering
\vspace{5mm}\includegraphics[width=0.5\textwidth]{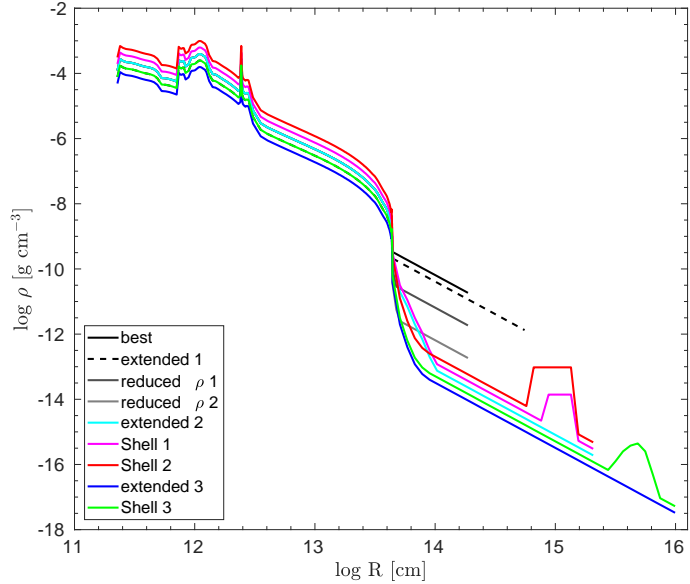}
\caption{Density structures of the models with CSM structures specified
in Table~\ref{table:addModels}. The profiles of the models ``extended\,1'' ($-0.2$),
``extended\,3'' ($-0.4$),  ``Shell\,1'' ($+0.2$), ``Shell\,2'' ($+0.4$), and
``Shell\,3'' ($-0.2$) are shown with different offsets (specified in parentheses)
for clarity.}
\label{figure:addRhoR}
\end{figure}

\begin{figure*}
\centering
\vspace{2mm}
\includegraphics[width=0.48\textwidth]{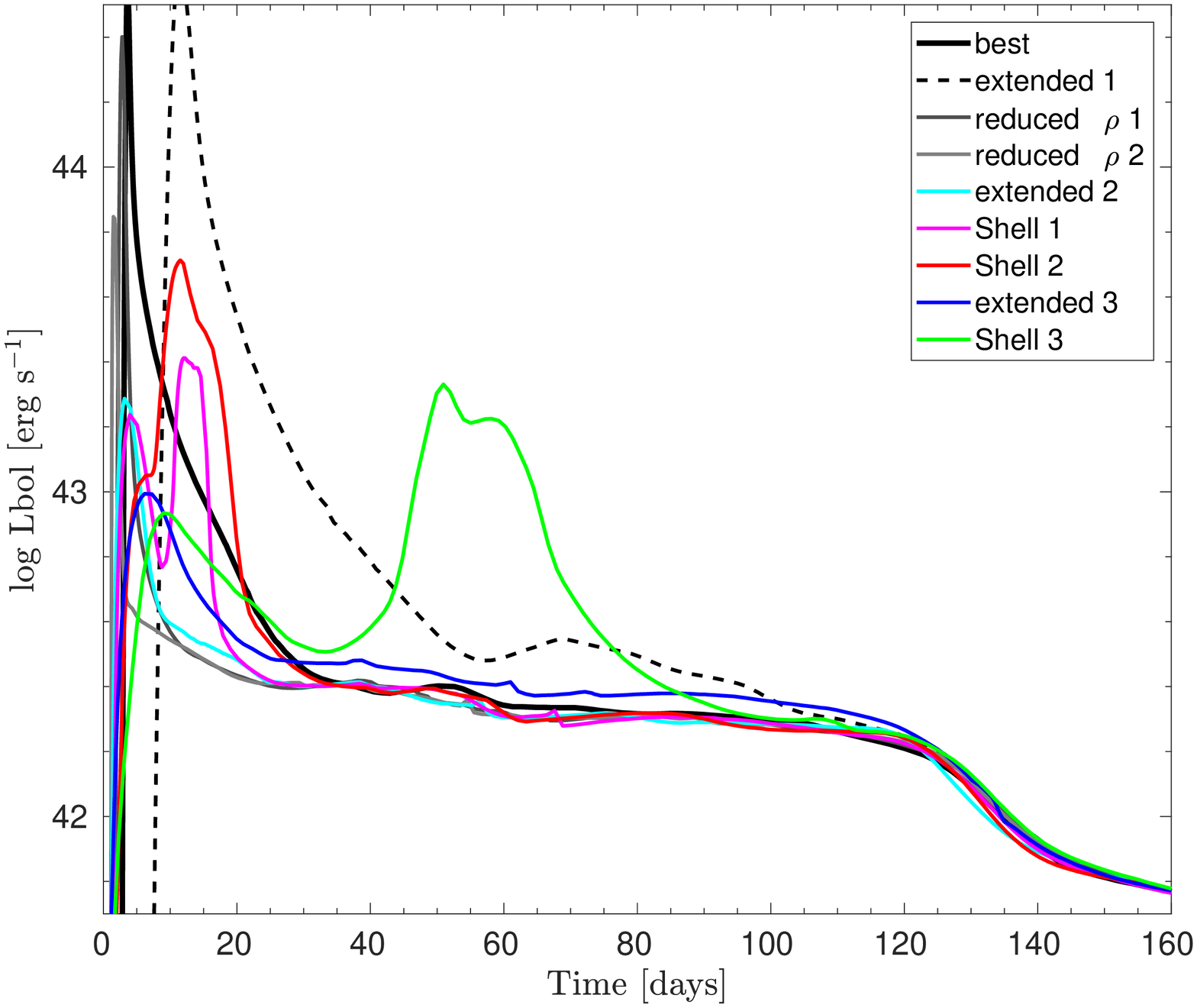}\hspace{4mm}
\includegraphics[width=0.48\textwidth]{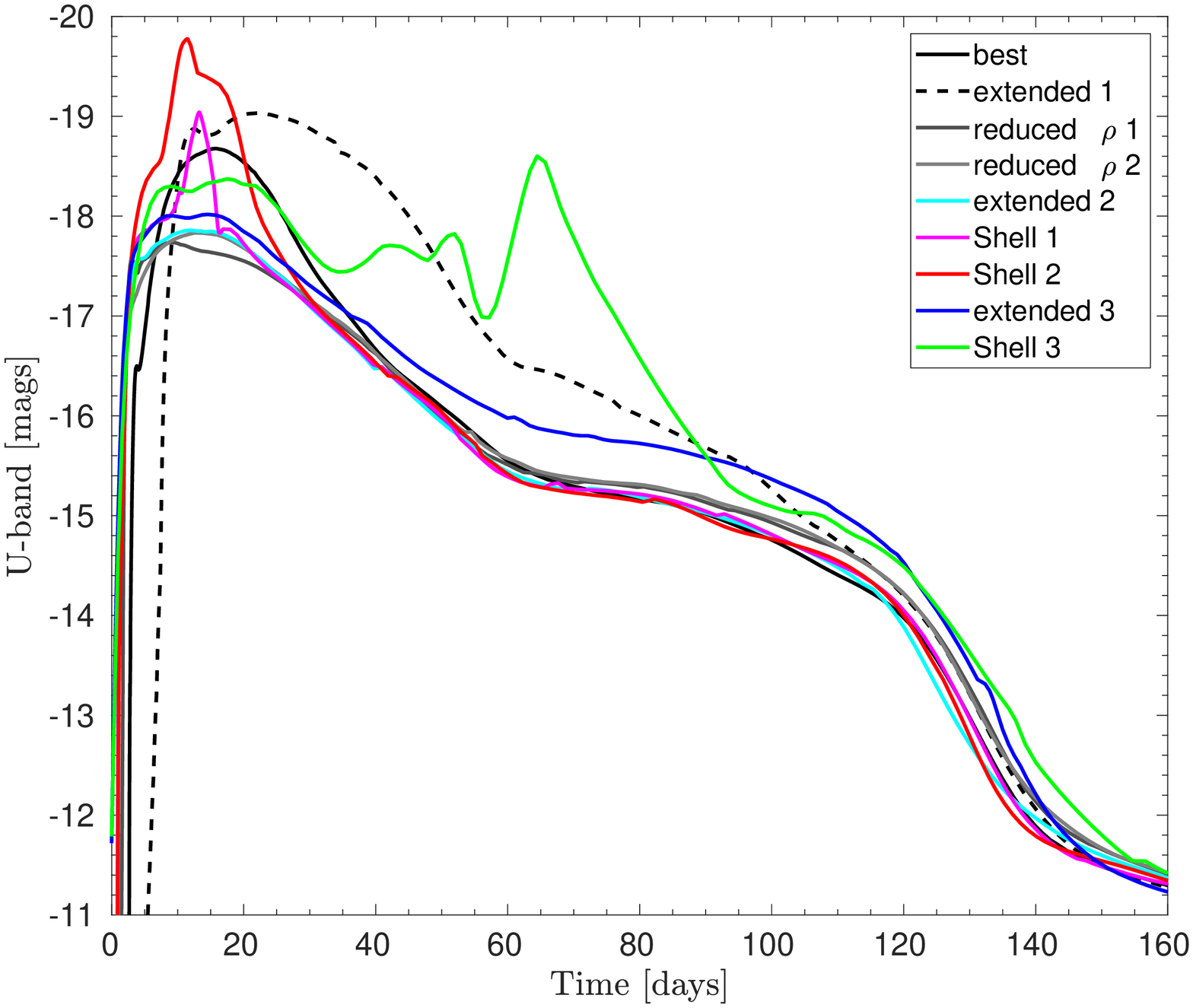}\\
\vspace{5mm}
\includegraphics[width=0.48\textwidth]{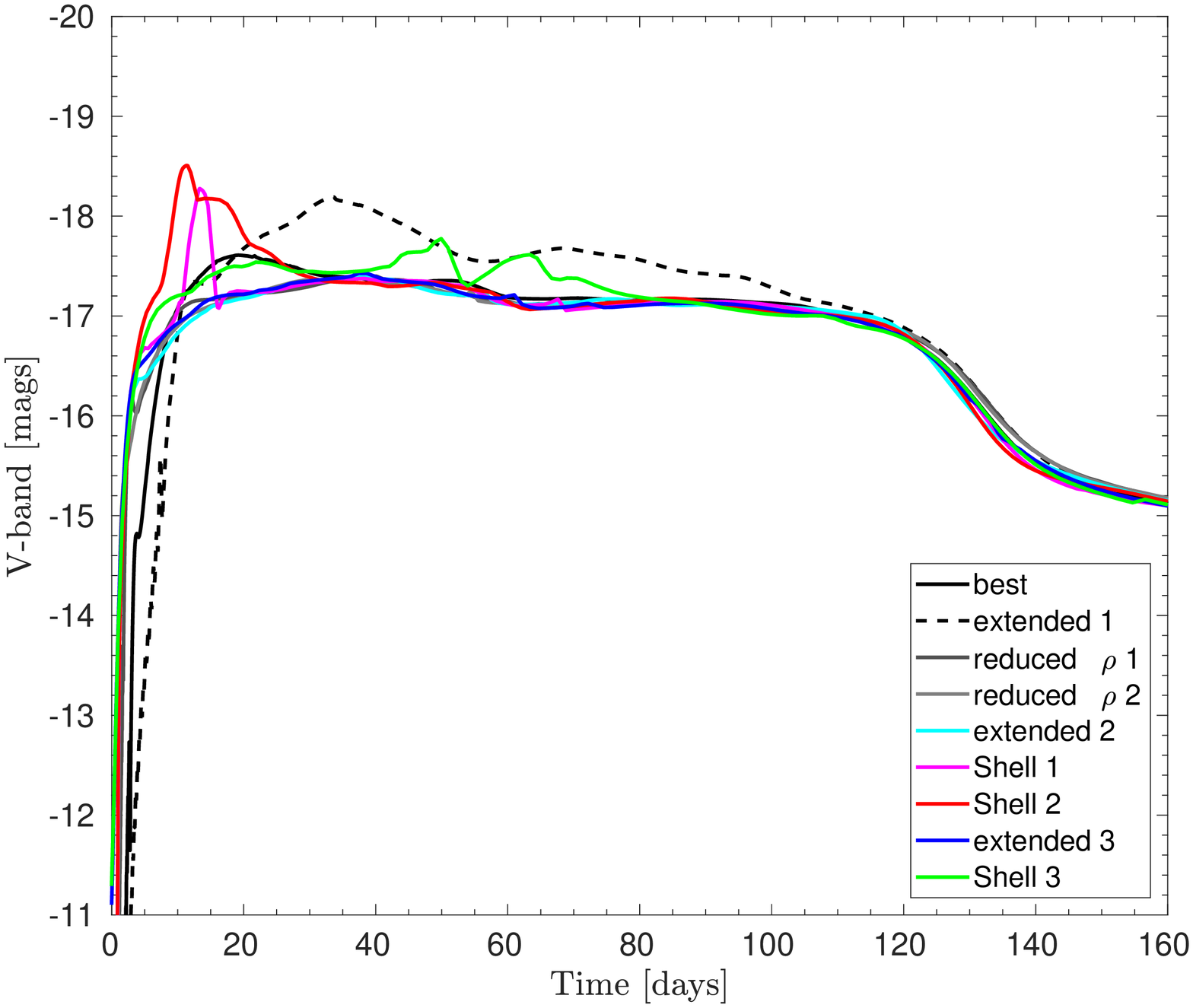}\hspace{4mm}
\includegraphics[width=0.48\textwidth]{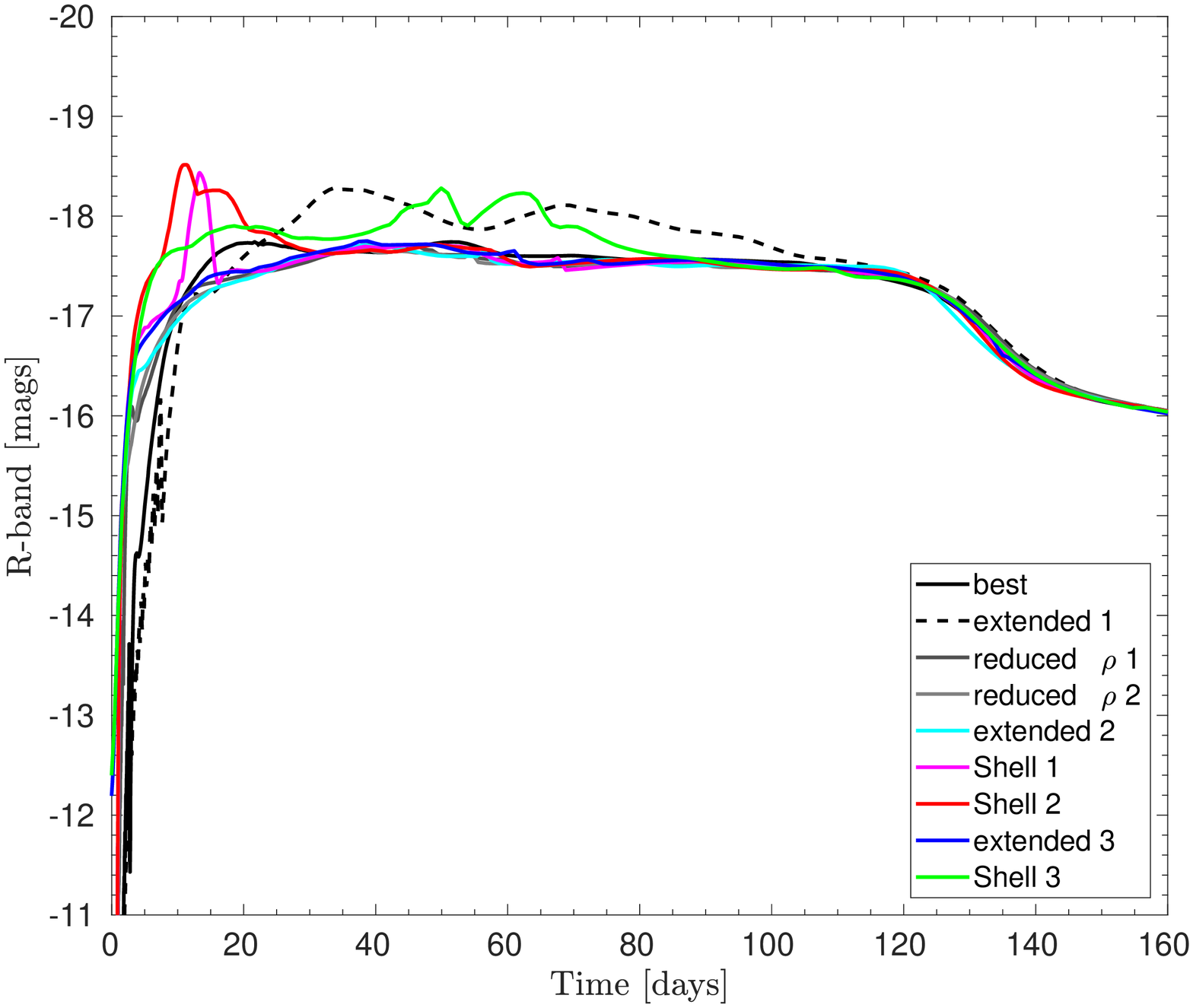}
\caption{Bolometric, $U$, $V$, and $R$ light curves for the models in
Table~\ref{table:addModels} and in Figure~\ref{figure:addRhoR}. The colours
represent the same models as in Figure~\ref{figure:addRhoR}.} 
\label{figure:addLCs}
\end{figure*}

The models ``reduced $\rho$ 1'' and ``reduced $\rho$ 2'' are shown in dark
grey and light grey in Figures~\ref{figure:addRhoR} and
\ref{figure:addLCs}. The lower density (factors of 10 and 100 lower than the
best-fit model) for the same extension of
CSM leads to a lower flux in the early-time bolometric light curves. For the ``reduced $\rho$ 2''
model the bolometric light curve is very close to the basic progenitor model without
CSM (see Fig.~\ref{figure:bol}).

The more extended model ``extended 1'' with the same ``interface'' density
(density where the CSM is attached to the progenitor) but larger outer radius (almost
three times larger) tend to gain higher mass
(1.95,\Msun\ vs. 0.55\,\Msun); consequently, the shock breakout in the CSM is
delayed and the bolometric light curve is broader (the dashed black lines in
Fig.~\ref{figure:addLCs}). The $U$-band light curve of the more extended model ``extended 1''
has a factor of two longer duration and is 0.5\,mag brighter. $V$-band magnitude
for the extended model  ``extended 1'' is brighter than the best-fit model,
and its plateau is affected during the entire duration by the presence of the additional matter.
The extended models ``extended 2'' and ``extended 3'' (represented by cyan
and blue in the figure) have the same ``interface'' density
(a factor of 1000 lower than in the best-fit model), and different extension
of 30,000\,\Rsun\ and 143,000\,\Rsun, respectively. Both ``extended 2'' and ``extended 3''
are affected only during the first 20\,days (see the bolometric light curves). For
the more extended case (``extended 3''),
the shock breaks at lower luminosity and a has slower decline,
consistent with the shock-cooling models. The broad-band light curves are
very similar to those of ``extended 1'' --- even the $U$-band magnitude, which is
usually the most sensitive to any changes in the density structure and
composition of the SN ejecta. 

The cases ``Shell 1,'' ``Shell 2,'' and ``Shell 3'' (magenta, red, and green) show very
distinct behaviour of their light curves. All three cases have two maxima; however,
``Shell 2'' has two maxima merged into one. The first
peak represents the shock breakout at the edge of the CSM including the shell, and
the second peak is the reproduced energy from the shock passage in the shell.
``Shell 1'' and ``Shell 2'' are inserted with the same CSM density
profile, although their masses differ significantly (0.03\,\Msun\ and
0.17\,\Msun, respectively). The first maximum in the ``Shell 2'' model has a lower
luminosity because of the larger mass involved. The second maxima in both ``Shell 1'' and ``Shell 2''
have similar shape (as the density structures have similar shape but differ in
amplitude). The larger the mass in the shell, the longer and the brighter
is the second maximum, which is the release of thermal energy after
the shock passage.
The model ``Shell 3'' has the shell localed at a larger distance;
consequently, both maxima are delayed. Surprisingly, the first maximum of the ``Shell 3''
case is similar in duration and luminosity to the best-fit model in broad bands,
even though the flux is distributed differently, and ``Shell 3'' has larger
red flux ($R$ band) and lower blue flux ($U$ band) than the ``best'' case.
The second maximum occurs at
significantly later times, 50\,days later than in the ``Shell 1'' and ``Shell 2''
cases. This is explained by the lower density (more than 10 times lower in``Shell 3'')
and larger radius (a factor of 10 larger in ``Shell 3''),
which slows the cooling processes in the ionised medium.
Maybe some configuration can be found to mimic the best-fit observables, but
it is beyond the scope of this paper.

%  To conclude, the presence of CSM in close vicinity (not beyond a
%few thousand solar radii) of the progenitor is
%required to reproduce the photometric properties of SN~2021yja. The mass of
%CSM could not be very different from 0.55\,\Msun\ because it would capture
%a different fraction of the shock energy and reprocess it differently.
%The smooth broad-band light curves of SN~2021yja indicate that the CSM
%has to be adjacent to the progenitor without gaps (i.e., no detached shells);
%otherwise, the light curve would have more than one maximum.

\begin{table*}
\caption{Physical properties of the CSM surrounding the best-fit progenitor model
used in the current study.}
\label{table:addModels}
\begin{tabular}{|l|c|c|c|c|}
\hline
& $\rho_0/\rho_\mathrm{out}$ [g\,cm$^{\,-3}$]& $R_\mathrm{out}$ [cm/\Rsun] &
$M_\mathrm{CSM}$ [\Msun]& $M_\mathrm{shell}$ [\Msun]\\
\hline
best       & $3\times10^{\,-10}/1.9\times10^{\,-11}$ & $1.9\times10^{\,14}$/2,700  & 0.55 & 0 \\
extended 1 & $3\times10^{\,-10}/2.1\times10^{\,-12}$ & $5.5\times10^{\,14}$/7,900  & 1.95 & 0 \\
reduced $\rho$ 1 & $3.8\times10^{\,-11}/1.8\times10^{\,-12}$ & $1.9\times10^{\,14}$/2,700 & 0.055 & 0 \\
reduced $\rho$ 2 & $3.8\times10^{\,-12}/1.8\times10^{\,-13}$ & $1.9\times10^{\,14}$/2,700 & 0.0056 & 0 \\
extended 2 & $3\times10^{\,-13}/1.9\times10^{\,-16}$ & $2\times10^{\,15}$/30,000  & 0.047  & 0 \\
Shell 1    & $3\times10^{\,-13}/1.9\times10^{\,-16}$ & $2\times10^{\,15}$/30,000  & 0.047  & 0.035 \\
Shell 2    & $3\times10^{\,-13}/1.9\times10^{\,-16}$ & $2\times10^{\,15}$/30,000  & 0.047  & 0.175 \\
extended 3 & $3\times10^{\,-13}/8.3\times10^{\,-18}$ & $1\times10^{\,16}$/143,000 & 0.045 & 0 \\
Shell 3    & $3\times10^{\,-13}/8.3\times10^{\,-18}$ & $1\times10^{\,16}$/143,000 & 0.045 & 0.16 \\
\hline
\end{tabular}
\tablecomments{The variables $\rho_0$ and $\rho_\mathrm{out}$ are
respectively the density where the CSM is attached to the progenitor and
the outer density, and $R_\mathrm{out}$ is the outer radius of the CSM.
$M_\mathrm{CSM}$ and $M_\mathrm{shell}$ are the mass of the CSM (without
the shell) and the mass of the shell separately, in solar masses. The model
``best'' is the successful model used to explain the photometric properties
of SN~2021yja.}
\end{table*}

\end{document}